\documentclass[preprint,12pt]{./elsarticle}
\usepackage{natbib}

\usepackage{amssymb}

\usepackage{amsfonts,amsmath,amssymb,amsbsy,amsthm}

\usepackage{hyperref}
\usepackage{mathrsfs}
\usepackage{epstopdf}

\usepackage{graphicx}

\usepackage{subcaption}

\usepackage{bbm} 

\begin{document}

\begin{frontmatter}

\title{A Critical Study of Efrati \emph{et al.}'s \emph{Elastic Theory of Unconstrained non-Euclidean Plates}}

\author{Kavinda Jayawardana\corref{mycorrespondingauthor}}
\cortext[mycorrespondingauthor]{Corresponding author}
\ead{zcahe58@ucl.ac.uk}

\begin{abstract}
In our analysis, we show that Efrati \emph{et al.}'s  publication \cite{efrati2009elastic} is inconsistent with the mathematics of plate theory. However it is more consistent with the mathematics of shell theory, but with an incorrect strain tensor. Thus, the authors' numerical results imply that a thin object can be stretched substantially with very little force, which is physically unrealistic and mathematically disprovable. All the theoretical work of the authors, i.e. nonlinear plate equations in curvilinear coordinates, can easily be rectified with the inclusion of both a sufficiently differentiable diffeomorphism and a set of external loadings, such as an external strain field.
\end{abstract}


\end{frontmatter}

\section{Introduction}
\label{S:1}

A \emph{plate} is a structural element with planar dimensions that are large compared  to its thickness. Thus, plate theories are derived  from the three-dimensional elastic theory by making suitable assumptions concerning the kinematics of deformation or the state stress through the thickness of the lamina, thereby reducing three-dimensional elasticity problem into a two-dimensional one.\\

Efrati \emph{et al.} \cite{efrati2009elastic} present a model, defined as the  \emph{elastic theory of unconstrained non-Euclidean plates}, for modelling deformation of thin objects. The main application of the authors' work is in the study of natural growth of tissue such as growth of leaves and other natural slender bodies. Some numerical results are present, which is based on an example of a hemispherical plate, and they imply the occurrence of buckling transition, from a stretching-dominated configuration to a bending-dominated configuration, under variation of the plate thickness.\\

However, we show that what the authors present is not a plate theory model; It is, in fact, a shell theory model, but with an incorrect strain tensor. Thus, the authors numerical results imply that a thin object can be stretched substantially with very little force, which is physically unrealistic and mathematically disprovable. All the theoretical work of the authors, i.e. nonlinear plate equations in curvilinear coordinates, can easily be rectified with the inclusion of both a sufficiently differentiable $\mathbb{R}^2 \to \textbf{E}^2$ diffeomorphism and a set of external loadings, such as an external strain field.

\section{Efrati \emph{et al.}'s Work}

Given that a growing leaf can be modelled by a plate, Efrati \emph{et al.} \cite{ efrati2009elastic}\footnote{\href{http://www.ma.huji.ac.il/~razk/iWeb/My_Site/Publications_files/ESK08.pdf}{ http://www.ma.huji.ac.il/$\sim$razk/iWeb/My\_Site/Publications\_files/ESK08.pdf}} focus on the elastic response of the plate after its planar (i.e. rest) configuration is modified, either by growth or by a plastic deformation. Thus, the goal is to derive a thin plate theory, as a generalisation of existing elastic plate theories, that it is valid for large displacements and small strains in arbitrary intrinsic geometries. The authors ignore the thermodynamic limitations on plastic deformations as they are considered to be not relevant when modelling naturally growing tissue, and further assume that the reference configuration is a known quantity. Their main postulate is that a non-Euclidean plate cannot assume a rest configuration, i.e. no stress-free configuration can exist, and thus, one faces a nontrivial problem that always exhibits residual stress. Note that by \emph{non-Euclidean} the authors mean `the internal geometry of the plate is not immersible in 3D Euclidean space' \cite{efrati2009elastic}. Also, the authors define `a metric is immersible in $\textbf{E}^3$' if Ricci curvature tensor with respect to the implicit coordinate system is identically zero, i.e. for a given immersion $\boldsymbol \varphi : \mathbb{R}^n \to \textbf{E}^3$, where $1\leq n \leq 3$, the metric $\bar{\boldsymbol g}$ on $\mathbb{R}^n$ induced by the immersion $\boldsymbol \varphi$ results in $\boldsymbol {Ric} = \boldsymbol 0$ in $\mathbb{R}^n$, where $\mathbb{R}^n$ and $\textbf{E}^n$ are curvilinear and Euclidean spaces respectively.\\

The authors define a plate as an elastic medium for which there exists a curvilinear set of coordinates $\boldsymbol x = \boldsymbol(x^1,x^2,x^3\boldsymbol)$, in which the `reference metric', $\bar g_{ij} = \bar g_{ij} (x^1,x^2)$, takes the form $\bar g_{\alpha \beta} = \bar g_{ \beta\alpha}$, $\bar g_{\alpha 3} = 0$, $\bar g_{33} = 1$. The reference metric is a symmetric positive-definite tensor and considered  to be a known quantity. The plate is considered to be `even', i.e. the domain $\mathscr{D} \subset \mathbb{R}^3$ of curvilinear coordinates can be decomposed into $\mathscr{D} = \mathscr{S} \times [-\frac{h}{2},\frac{h}{2}]$, where $\mathscr{S} \subset \mathbb{R}^2$ and $h$ is the thickness of the plate. Thus, it is given that an even plate is fully characterised by the metric of its mid-surface, i.e. at where $x^3= 0$.\\

Although thin plates are three-dimensional bodies, the authors took advantage of the large aspect ratio by modelling the plates as two-dimensional surfaces, and thus, reducing the dimensionality of the problem. To achieve this the authors assume `Kirchhoff-Love assumptions:' (i) `the body is in a state of plane-stress (the stress is parallel to the deformed mid-surface)', and (ii) `points which are located in the undeformed configuration on the normal to the mid-surface at a point $p$, remain in the deformed state on the normal to the mid-surface at $p$, and their distance to $p$ remains unchanged' \cite{efrati2009elastic}.\\

Now, consider the deformed plate in the Euclidean space, which is defined as a compact domain $\Omega \subset \mathbb{R}^3$ endowed with a regular set of material curvilinear coordinates. Define the mapping $\boldsymbol r: \mathscr{D}\subset \mathbb{R}^3 \mapsto \Omega$, from the domain of parameterisation $\mathscr{D}$ into $\Omega$, as the configuration of the body endowed with the metric tensor $\boldsymbol g$, which is defined as $g_{ij} = \partial_i \boldsymbol r \cdot \partial_j \boldsymbol r$, where $\cdot$ is the Euclidean dot-product. It is given that Kirchhoff-Love second assumption implies that $g_{\alpha 3} = 0$. Thus, when defined more precisely, one finds that $\boldsymbol r( x^1,x^2,x^3) = \boldsymbol R( x^1,x^2) + x^3\hat{\boldsymbol N}(x^1,x^2)$, $g_{\alpha\beta} = a_{\alpha\beta} - 2x^3b_{\alpha\beta} + (x^3)^2 c_{\alpha\beta}$, $g_{\alpha 3} = 0$ and $g_{33} = 1$, where $\boldsymbol R$ is the mid-surface, $\hat{\boldsymbol N}$ is the unit normal to the mid-surface, and $a_{\alpha\beta}$, $b_{\alpha\beta}$ and $c_{\alpha\beta}$ are the first, the second and the third fundamental form tensors respectively. With further inspection one finds that $a_{\alpha\beta} = \partial_\alpha \boldsymbol R \cdot \partial_\beta \boldsymbol R$, $b_{\alpha\beta} = (\partial_\alpha \partial_\beta \boldsymbol R) \cdot \hat{\boldsymbol N}$ and $c_{\alpha\beta} = (a^{-1})^{\gamma \delta}b_{\alpha\gamma} b_{\beta \delta}$. The ultimate goal is to find the metric tensor $\boldsymbol g$, and the authors state that the metric tensor $\boldsymbol g$ is immersed in $\mathbb{R}^3$, and thus, the metric tensor uniquely defines the physical configuration of a three-dimensional body. It is also the case that one needs to find equations to six unknowns which make up the metric tensor for the general case, where $\bar{\boldsymbol g}$ is not defined by $\boldsymbol r$. For the general case, the authors describe one approach to this problem via the use of `the modified version of the hyper-elasticity principle ... the elastic energy stored within a deformed elastic body can be written as a volume integral of a local elastic energy density, which depends only on (i) the local value of the metric tensor and (ii) local material properties that are independent of the configuration' \cite{efrati2009elastic}. It is unclear what the authors mean by this definition; thus, for a more precise definition of hyperelasticity, we refer the reader to Ball \cite{Ball} or Ciarlet \cite{Ciarlet}.\\

The authors define the strain tensor as follows,
\begin{align}
\epsilon_{ij} & = \frac{1}{2} (g_{ij} - \bar g_{ij})~, \label{shStrain}
\end{align}
and thus, the energy functional is expressed as follows,
\begin{align}
E(g) & =\int_{\mathscr{D}} w(g) \sqrt{\bar g}~dx^1 dx^2 dx^3 ~,\label{sh1}
\end{align}
where $w = \frac{1}{2} A^{ijkl} \epsilon_{ij} \epsilon_{kl}$ is the energy density and $A^{ijkl} = \lambda \bar g^{ij} \bar g^{kl} + \mu (\bar g^{ik} \bar g^{jl} + \bar g^{il} \bar g^{jk})$ is the elasticity tensor. With the use of the energy functional (\ref{sh1}), the configuration $\boldsymbol r$ is varied to find the three constraints that $g_{\alpha\beta}$ must satisfy, i.e. `the fundamental model for three-dimensional elasticity' \cite{efrati2009elastic}. 
Note that \emph{Einstein's summation notation} is assumed  throughout, and we regard the indices $i,j,k,l \in \{1,2,3\}$ and  $\alpha,\beta,\gamma,\delta \in \{1,2\}$, unless it is strictly states otherwise. Also note that the authors define the symmetric Ricci curvature tensor of the metric $\boldsymbol g$ as follows,
\begin{align*}
Ric_{li}  = \frac{1}{2} \left(g^{-1})^{kj}(\partial_k\partial_i g_{lj} - \partial_k \partial_j g_{li} + \partial_j \partial_l g_{ki} - \partial_i \partial_l g_{kj}\right)&\\
+ (g^{-1})^{kj} g_{pq} (\Gamma^p_{\!lj}\Gamma^q_{\!ki} - \Gamma^p_{\!kj}\Gamma^q_{\!li}) &~.
\end{align*}

As the elastic body is immersed in $\mathbb{R}^3$,  the variational principle implies that the six independent components of the symmetric Ricci curvature tensor must all vanish, i.e. $Ric_{ij} = 0$. However, $Ric_{ij} = 0$ and the three equations obtained  by varying the configuration $\boldsymbol r$ in equation (\ref{sh1}) imply that the system is over-determined. Thus, the authors postulate that there are two possible ways to resolve this `seemingly over-determination'. The first is by noticing that the six independent components of Ricci curvature tensor's derivatives are related through second Bianchi identity. The second way of resolving this issue is by identifying the immersion $\boldsymbol r$ as the three unknown functions (as defined previously), in which case the six equations that form the Ricci tensor are the solvability conditions for the partial differential equation. However, as the equations in $\boldsymbol r$ are of the higher order, one needs to supply additional conditions, namely to set the position and the orientation of the body, in order to obtain a unique solution for $\boldsymbol r$.\\

To find the reduced  energy density the authors integrate the energy density (\ref{sh1}) over the thin dimension as $w_{2D} = \int^{\frac{h}{2}}_{-\frac{h}{2}} w ~dx^3$ to obtain the  equation,
\begin{align*}
w_{2D} & = h w_S + h^3 w_B~,
\end{align*}
where
\begin{align}
w_S & = \frac{Y}{8(1+\nu)} \left(\frac{\nu}{1-\nu}\bar g^{\alpha \beta} \bar g^{\gamma \delta} +\bar g^{\alpha \gamma}\bar g^{\beta \delta}\right)(a_{\alpha \beta} - \bar g_{\alpha \beta})(a_{\gamma \delta} - \bar g_{\gamma \delta}) ~~\text{and}\nonumber\\
w_B & = \frac{Y}{24(1+\nu)}\left(\frac{\nu}{1-\nu}\bar g^{\alpha \beta} \bar g^{\gamma \delta} +\bar g^{\alpha \gamma}\bar g^{\beta \delta}\right)b_{\alpha \beta}b_{\gamma \delta} ~, \label{efratiWrong}
\end{align}
which are defined as stretching and bending densities respectively. Note that $Y$ is the Young's modulus and $\nu$ is the Poisson's ratio of the plate.\\

It is stated that with the use of Cayley-Hamilton theorem, the density of the bending content can be written in the following form,
\begin{align*}
w_B = \frac{Y}{8(1+\nu)} \big(\frac{1}{1-\nu}(\bar g^{\alpha \beta} b_{\alpha \beta})^2 - 2 \frac{|b|}{|\bar g|}\big) ~.
\end{align*}

It is also stated that, if $a_{\alpha \beta} = \bar g_{\alpha \beta}$ (i.e. the two-dimensional configuration has zero-stretching energy), then the density of the bending content can be expressed as the density of  Willmore functional \cite{willmore1996riemannian} as follows,
\begin{align*}
w_W & = \frac{Y}{24(1+\nu)}\left(\frac{4H^2}{1-\nu} - 2K\right) ~,
\end{align*}
where $K$ and $H$ are Gaussian and the mean curvatures of the mid-surface respectively.\\

With the vanishing of the Ricci tensor the authors obtain Gaussian curvature and Gauss-Mainardi-Peterson-Codazzi equations, which are  respectively defined as
\begin{align}
K = \frac{|b|}{|a|} & =\frac{1}{2}(a^{-1})^{\alpha\beta}(\partial_\gamma \Gamma^\gamma_{\!\alpha \beta} - \partial_\beta \Gamma^\gamma_{\!\alpha \gamma} + \Gamma^\gamma_{\!\gamma\delta}\Gamma^\delta_{\!\alpha\beta} - \Gamma^{\gamma}_{\!\beta\delta}\Gamma^\delta_{\!\alpha\gamma}) ~~\text{and}\label{shar1}\\
\partial_2b_{\alpha 1} + \Gamma^\beta_{\!\alpha 1} b_{\beta 2} & = \partial_1 b_{\alpha 2} + \Gamma^\beta_{\!\alpha 2} b_{\beta 1}~, \label{shar2}
\end{align}
where equations (\ref{shar1}) and (\ref{shar2}) given to provide sufficient conditions for immersiblility of the metric tensor $\boldsymbol g$ in $\mathbb{R}^3$.\\

It may appear to the uninitiated in the study mathematical elasticity that Efrati \emph{et al.}'s  publication \cite{efrati2009elastic} is a coherent piece of work, but it is, in fact, flawed. To illustrate this matter in detail, we direct the reader's attention to section 3.4 of Efrati \emph{et al.} \cite{efrati2009elastic}. Upon examining the governing equations and the boundary conditions, one can see that the governing equations are defined for zero-external loadings, and the boundary conditions are defined for zero-tractions, zero-boundary moments and there are no descriptions of any Dirichlet conditions. Thus, it is mathematically impossible to obtain a non-zero solution (excluding any rigid motions). Furthermore, there is no evidence in the authors' publication for proof of the existence of solutions, either via rigorous mathematics ($\Gamma$-limit or otherwise) or via numerical analysis (only some conjectures regarding solvability conditions are given by the authors, as we previously discussed).\\

\begin{figure}[!h]
\centering
\includegraphics[width=1\linewidth]{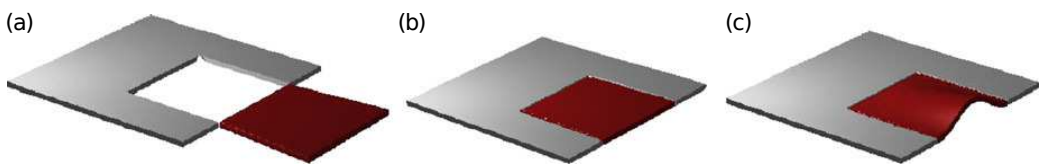}
\caption{`A schematic illustration of an unconstrained  plate exhibiting residual stress. (a) The two elements composing the plate are shows side by side. (b) As the red trapezoid is too large to fit into the square opening, it is compressed. (c) For a plate sufficiently thin, the induced compression exceeds the buckling threshold, and the trapezoid buckles out of plane. Note that there are many shapes that preserve all lengths along the faces of the plate, yet they cannot be planar'  \cite{efrati2009elastic}.\label{EfratiPic01}}
\end{figure}

Efrati \emph{et al.}'s \cite{efrati2009elastic} erroneous work arises from not fully understanding how to model the given problem. Consider figure \ref{EfratiPic01} (c) (see figure 1 from Efrati \emph{et al.} \cite{efrati2009elastic}): the very reason the \emph{red trapezoid} is deformed is because it is compressed at the boundary, i.e. it is deformed as it is subjected to a Dirichlet boundary condition. Thus, if one attempts to model this problem with mathematical rigour, then one can derive the actual energy functional for this problem. To do so, consider the  map
\begin{align}
\boldsymbol r (\boldsymbol u) & = \boldsymbol R (\boldsymbol u)+ z \frac{\partial_x \boldsymbol R(\boldsymbol u) \times \partial_y \boldsymbol R(\boldsymbol u)}{|| \partial_x \boldsymbol R (\boldsymbol u) \times \partial_y \boldsymbol R(\boldsymbol u)||} ~,\label{mapEfrati}
\end{align}
which is assumed to be a sufficiently differentiable $\textbf{E}^3 \to \textbf{E}^3$ diffeomorphism for an appropriate $\boldsymbol u \in \textbf{E}^3$, where $$\boldsymbol R(\boldsymbol u) = \boldsymbol ( x, ~y, ~0 \boldsymbol)_\text{E} + \boldsymbol(u^{\bar1}(x,y), ~u^{\bar2}(x,y), ~u^{\bar3}(x,y) \boldsymbol)_\text{E}$$ is the deformed mid-surface of the plate, $\boldsymbol u$ is the displacement field that describes a vector displacement in the three-dimensional Euclidean space and $\times$ is the Euclidean cross-product. The metric with respect to map (\ref{mapEfrati}) in the Euclidean space is $g_{\bar\alpha \bar\beta}(\boldsymbol u) = \partial_{\bar\alpha} r_{\bar i }(\boldsymbol u)\partial_{\bar\beta} r^{\bar i} (\boldsymbol u)$, where the over-bar in the indices highlights the fact that one is using Euclidean coordinates. Now, the strain tensor of a plate can be expressed as follows
\begin{align*}
\epsilon_{\bar\alpha \bar\beta}(\boldsymbol u) & = \frac{1}{2}\left (g_{\bar\alpha \bar\beta} (\boldsymbol u) - \delta_{\bar\alpha \bar\beta}\right)~,
\end{align*}
where $x^{\bar1}=x$ and $x^{\bar2}=y$. Thus, the energy functional can be expressed as follows,
\begin{align}
J(\boldsymbol u) = \int\limits_\mathscr{S} \Bigg[\frac{1}{8} h A^{\bar\alpha \bar\beta \bar\gamma \bar\delta} & \left(\partial_{\bar\alpha} R_{\bar i}(\boldsymbol u) \partial_{\bar \beta} R^{\bar i}(\boldsymbol u) -\delta_{\bar \alpha \bar\beta}\right) \left(\partial_{\bar \gamma} R_{\bar k}(\boldsymbol u) \partial_{\bar\delta} R^{\bar k}(\boldsymbol u) -\delta_{\bar\gamma\bar\delta}\right) \nonumber \\
+\frac{1}{24}h^3A^{\bar \alpha \bar \beta \bar \gamma \bar \delta} & \left(\partial_{\bar \alpha \bar \beta} R_{\bar i}(\boldsymbol u) \frac{\left(\partial_x \boldsymbol R(\boldsymbol u) \times \partial_y \boldsymbol R(\boldsymbol u)\right)^{\bar i}}{|| \partial_x \boldsymbol R (\boldsymbol u) \times \partial_y \boldsymbol R(\boldsymbol u)||}\right ) \nonumber \\
 & \left(\partial_{\bar \gamma \bar \delta} R_{\bar k}(\boldsymbol u) \frac{\left(\partial_x \boldsymbol R(\boldsymbol u) \times \partial_y \boldsymbol R(\boldsymbol u)\right)^{\bar k}}{|| \partial_x \boldsymbol R (\boldsymbol u) \times \partial_y \boldsymbol R(\boldsymbol u)||}\right )  \Bigg]dx dy~, \label{nonlinerPlate}\\
\boldsymbol u  \in & \{ \boldsymbol v \in \boldsymbol W (\mathscr{S}) \mid \boldsymbol v|_{\partial\mathscr{S}} = \boldsymbol u_0\}~,\label{nolinearSet}
\end{align}
where $\mathscr{S}\subset \textbf{E}^2$ is the mid-plane of the unstrained plate, $ \boldsymbol W (\mathscr{S})$ is an appropriate Sobolev space, $\boldsymbol u_0 \in L^2(\partial\mathscr{S})$ is a Dirichlet boundary condition, and $A^{\bar \alpha \beta \bar \gamma \bar \delta} = \frac{1}{2} (1+\nu)^{-1}Y(2(1-\nu)^{-1} \nu\delta^{\bar \alpha \bar \beta}\delta^{\bar \gamma \bar \delta} + \delta^{\bar \alpha\bar \gamma}\delta^{\bar \beta \bar \delta} + \delta^{\bar \alpha \bar \delta}\delta^{\bar \beta \bar\gamma})$ is the elasticity tensor of a plate. As $\boldsymbol R (\boldsymbol u)$ describes a surface, one may argue that with an appropriate coordinate transform, one can express equation (\ref{nonlinerPlate}) in the same form as the authors' energy functional (see equation 3.7 of Efrati \emph{et al.} \cite{ efrati2009elastic}), which is  true if one is using an appropriate $\boldsymbol(x,y\boldsymbol)_\text{E}: \mathbb{R}^2 \to \textbf{E}^2$ coordinate transform (the authors' erroneous coordinate transform leads to a reference metric of a shell at $x^3=0$; please see equation 4.1 and see the definition of $\Phi(r)$ from section 4.2 of Efrati \emph{et al.} \cite{ efrati2009elastic}). However, without the Dirichlet boundary condition from equation (\ref{nolinearSet}) or some other external loading, which is exactly what the authors are considering (see respectively equation 2.5 and section 3.2 of Efrati \emph{et al.} \cite{ efrati2009elastic}), one gets the trivial zero-displacement solution, i.e. $\boldsymbol u =\boldsymbol 0$ in $\mathscr{S}$. The fact that the authors are claiming that nonzero solutions are possible without tractions, Dirichlet boundary conditions, external loadings or boundary moments imply that they are doing something fundamentally flawed.\\

Now, consider the sufficiently differentiable diffeomorphism $\boldsymbol \varphi : \omega \subset \mathbb{R}^2 \to \boldsymbol \varphi (\omega ) \subset \textbf{E}^2$ with the property $\det(\boldsymbol J)>0$, where $\boldsymbol(x,y\boldsymbol)_\text{E} = \boldsymbol(\varphi^{\bar1}(x^1,x^2),$ $\varphi^{\bar2}(x^1,x^2)\boldsymbol)_\text{E} $ and $J^{\bar \beta}_\alpha = \partial_\alpha \varphi ^{\bar \beta}$ is the Jacobian matrix of the map $\boldsymbol \varphi $. Note that we reserve the vector brackets $\boldsymbol (\cdot \boldsymbol )_{\text{E}}$ for vectors in the Euclidean space and $\boldsymbol (\cdot \boldsymbol )$ for vectors in the curvilinear space. Now, consider the mapping $\boldsymbol R(\boldsymbol u)\circ \boldsymbol \varphi :\omega \subset \mathbb{R}^2 \to \boldsymbol R(\boldsymbol u)(\mathscr{S}) \subset \textbf{E}^3$. As $\boldsymbol \varphi$ is a diffeomorphism, $\boldsymbol R(\boldsymbol u)\circ \boldsymbol \varphi$ is a well defined surface for a suitable displacement field $\boldsymbol u \in \boldsymbol W (\mathscr{S})$, and as $\det(\boldsymbol J)>0$, the unit normal to the surface $\boldsymbol R(\boldsymbol u)$ is equal to the unit normal to the surface $\boldsymbol R(\boldsymbol u)\circ \boldsymbol \varphi$. Thus, with respect to $\boldsymbol \varphi$, equation (\ref{nonlinerPlate}) reduces to the following,
\begin{align}
J(\boldsymbol u)  =  \int\limits_\omega \Bigg[\frac{1}{2} A^{\alpha \beta \gamma \delta}\left( h a_{\alpha \beta} (\boldsymbol u)a_{\gamma \delta}(\boldsymbol u)  + \frac{1}{12} h^3 b_{\alpha \beta}(\boldsymbol u) b_{\gamma \delta}(\boldsymbol u)\right) &\nonumber\\
\sqrt{\det(\bar g)} &\Bigg] dx^1 dx^2 ~,\label{nonlinerPlate2}\\
\boldsymbol u  \in  \{ \boldsymbol v \in \boldsymbol W (\mathscr{S}) \mid \boldsymbol v|_{\partial\mathscr{S}} = \boldsymbol u_0\}&~,\nonumber
\end{align}
where
\begin{align*} 
a_{\alpha \beta} (\boldsymbol u) & = \frac12 \left((\partial_\alpha R_{\bar i}(\boldsymbol u)\circ \boldsymbol \varphi)( \partial_\beta R^{\bar i}(\boldsymbol u) \circ \boldsymbol \varphi)-\bar g_{\alpha\beta}\right) ~,\\
b_{\alpha \beta} (\boldsymbol u) & = (\partial_{ \alpha \beta} R_{\bar i}(\boldsymbol u) \circ \boldsymbol \varphi) \frac{\left((\partial_1 \boldsymbol R(\boldsymbol u)\circ \boldsymbol \varphi )\times (\partial_2 \boldsymbol R(\boldsymbol u)\circ \boldsymbol \varphi)\right)^{\bar i}}{|| (\partial_1 \boldsymbol R (\boldsymbol u)\circ \boldsymbol \varphi) \times (\partial_2 \boldsymbol R(\boldsymbol u)\circ \boldsymbol \varphi)||}~,\\
\bar g_{\alpha\beta} & = \partial_\alpha \varphi^{\bar\gamma} \partial_\beta \varphi_{\bar\gamma} ~,\\
A^{\alpha \beta \gamma \delta} & =  \frac{Y}{2(1+\nu)}\left( \frac{2\nu}{1-\nu} \bar g ^{ \alpha \beta}\bar g ^{\gamma \delta} + \bar g ^{\alpha\gamma}\bar g ^{\beta \delta} + \bar g ^{ \alpha \delta}\bar g ^{ \beta \gamma}\right )~.
\end{align*}
Now, equation (\ref{nonlinerPlate2}) is exactly the same form as the nonlinear plates equations in curvilinear coordinates put forward by the authors, excluding the Dirichlet boundary condition. However, equation (\ref{nonlinerPlate2}) is derived from the plate equations in Euclidean coordinates (\ref{nonlinerPlate}) with the use of the map $\boldsymbol \varphi$ which is a $\mathbb{R}^2 \to \textbf{E}^2 $ diffeomorphism, and thus, the reference metric $\bar{\boldsymbol g}$ is immersible in $\textbf{E}^2$. To be more precise, consider  Ricci curvature tensor in the two-dimensional Euclidean space, $Ric^\text{Euclidean}_{\bar \alpha \bar \beta} = Ric_{ \gamma \delta }(J^{-1})^\gamma_{\bar\alpha}(J^{-1})^\delta_{\bar\beta}$. As Ricci curvature tensor is identically zero in the Euclidean space (clearly!) and the map $\boldsymbol \varphi$ is a diffeomorphism, we have the vanishing of Ricci curvature tensor in $\omega\subset \mathbb{R}^2$, i.e. $\boldsymbol {Ric} =\boldsymbol 0$. Now, consider the map $\boldsymbol \varphi \times \boldsymbol(z \boldsymbol)_\text{E} :\{\omega\times[-\frac12,\frac12]\} \subset \mathbb{R}^2 \times\textbf{E} \to \{\boldsymbol \varphi (\omega)\times[-\frac12,\frac12]\} \subset \textbf{E}^3 $. As $\boldsymbol \varphi $ is a diffeomorphism, the map $\boldsymbol \varphi \times \boldsymbol(z \boldsymbol)_\text{E}$ is also a diffeomorphism, and thus, the metric generated by the map $\boldsymbol \varphi \times \boldsymbol(z \boldsymbol)_\text{E}$ is immersible in $\textbf{E}^3$. Furthermore, as $\boldsymbol r (\boldsymbol u) :\mathscr{S}\times[-\frac12,\frac12]\subset \textbf{E}^3 \to \boldsymbol r(\boldsymbol u)(\mathscr{S}\times[-\frac12,\frac12]) \subset \textbf{E}^3 $ is a diffeomorphism, the metric on $\textbf{E}^3$ induced by $\boldsymbol r(\boldsymbol u)$ (i.e. the metric of deformation) is immersible in $\textbf{E}^3$. Thus, the metric on $\mathbb{R}^2\!\times\!\textbf{E}$ generated by the map $r(\boldsymbol u)\circ\boldsymbol(\boldsymbol \varphi \times \boldsymbol(z \boldsymbol)_\text{E}\boldsymbol) :\{\omega\times[-\frac12,\frac12]\} \subset \mathbb{R}^2\!\times\!\textbf{E} \to \boldsymbol r(\boldsymbol u)(\mathscr{S}\times[-\frac12,\frac12]) \subset \textbf{E}^3 $ (i.e. the metric $\boldsymbol g$) is also immersible in $\textbf{E}^3$. However, the authors assert that their reference metric is not immersible in $\textbf{E}^3$. But it is mathematically impossible to derive equation (\ref{nonlinerPlate2}) from equation (\ref{nonlinerPlate}) without the use of a sufficiently differentiable $\mathbb{R}^2 \to \textbf{E}^2 $ diffeomorphism; thus, the fact that the authors claiming that their reference metric is not immersible in the three-dimensional Euclidean space (while their metric $\boldsymbol g$ is immersible in $\textbf{E}^3 $) means that the authors are attempting something fundamentally flawed.\\

Note that, if equation (\ref{nonlinerPlate}) linearised and along with the Dirichlet boundary condition, then one gets the following,
\begin{align*}
J(\boldsymbol u)  = \int\limits_\mathscr{S} \frac{1}{2} A^{\bar \alpha \bar \beta \bar \gamma \bar \delta}\left( h a_{\bar \alpha \bar \beta} (\boldsymbol u)a_{\bar \gamma \bar \delta}(\boldsymbol u) + \frac{1}{12} h^3 b_{\bar \alpha \bar \beta}(\boldsymbol u) b_{\bar \gamma \bar \delta}(\boldsymbol u)\right)dx dy & ~,\\
\boldsymbol u \in \{ \boldsymbol v \in H^1(\mathscr{S})\!\times\!H^1(\mathscr{S})\!\times\!H^2(\mathscr{S}) \mid \boldsymbol v|_{\partial\mathscr{S}_0} = \boldsymbol 0,&\\
~n^{\bar \alpha} \partial_{\bar \alpha} v^3|_{\partial\mathscr{S}_0} =0,~\boldsymbol v|_{\{\partial\mathscr{S}\setminus\partial\mathscr{S}_0\}} = \boldsymbol u_0&\}~,
\end{align*}
where $a_{\bar\alpha \bar \beta}(\boldsymbol u) = \frac{1}{2}(\partial_{\bar\alpha} u_{\bar\beta} + \partial_{ \bar\beta} u_{\bar\alpha})$, $b_{\bar\alpha \bar\beta} (\boldsymbol u) = \partial_{\bar\alpha\bar\beta}u^{\bar 3}$, $\boldsymbol n$ is the unit outward normal to the boundary $\partial \mathscr{S}$ and $\partial\mathscr{S}_0 \subset \partial\mathscr{S}$ with $\mathrm{meas}(\partial\mathscr{S}_0;\mathbb R)>0$, where $\mathrm{meas}(\cdot;\mathbb{R}^n)$ is the standard Lebesgue measure in $\mathbb{R}^n$ and $H^n(\cdot)$ are the standard $W^{n,2}(\cdot)$-Sobolev spaces (see section 5.2.1 of Evans \cite{Evans}). Such problems can be solved by consulting the literature that are specialised in the study of linear plate theory (see Ciarlet \cite{Ciarlet1997theory} and Reddy \cite{reddy2006theory}). \\

For numerical results, instead of finding the unknown metric tensor $\boldsymbol g$ (which is the goal of the publication), the authors attempt to analyse the stretching and the bending densities, i.e.  $w_s$ and $w_b$ respectively, for a predetermined reference metric $\bar{\boldsymbol g}$ and a predetermined deformed mid-surface $\boldsymbol R$, and thus, a predetermined deformed metric $\boldsymbol g$ (see section 4 of Efrati \emph{et al.} \cite{efrati2009elastic}). The authors give numerical results for an `annular hemispherical plate', i.e. annular plate deformed in to a hemispherical shape, and state that numerical results demonstrate that in the general case there is no `equipartition' between bending and stretching energies. The authors conclude by saying their numerical findings support treating very thin bodies as inextensible, and `it also shows that not only in the equilibrium 3D configuration dominated by the minimisation of the bending energy term, but the total elastic energy is dominated by it also' \cite{efrati2009elastic}. The reader must understand that the authors' numerical results do not imply the existence of a solutions (i.e. the existence of the deformation metric $\boldsymbol g$ not proven), as the numerical results are obtained for a predetermined metric $\boldsymbol g$.\\

The authors' numerical analysis implies that a thin object can be stretched substantially with very little force. To examine this in more detail, consider the following simple example in accordance with the authors' numerical analysis. Consider two circular plates: plate $c$ and plate $s$, with same Young's modulus $Y$, Poisson's ratio $\nu$, thickness $h$ and radius $r$, and assume that $h/r =\varepsilon \ll 1 $. Now, take plate $c$ and deform it into the shape of a semi-cylinder with a radius $\frac{2}{\pi} r$ (an area preserving deformation). Following the authors' publication, one finds that the mid-surface can be express by the following map, $$\boldsymbol R (x^1,x^2) = \boldsymbol(x^1, ~\frac{2}{\pi}r\sin(x^2), ~\frac{2}{\pi}r\cos(x^2)\boldsymbol)_\text{E}~,$$ and as one knows the deformed configuration in advance, one finds that the reference metric has the form  $\bar{\boldsymbol g} (x^1,x^2) = \mathrm{diag}(1, ~(\frac{2}{\pi}r)^2,~1)$. Thus, the stored energy of a circular plate that is being deformed into a semi-cylindrical shape can be expressed as follows,
\begin{align}
E_c &=\frac12\frac{Y}{1-\nu^2}\int_{-\frac{1}{2}h}^{\frac{1}{2}h}\iint_{\{(x^1)^2+(\frac{2r}{\pi}x^2)^2\leq r^2\}} \left(\frac{\pi x^3}{2r}\right)^2 \left(\frac{2r}{\pi}\right) dx^1 dx^2 dx^3 \nonumber\\ 
& = \frac{1}{96}\pi^3 \frac{Y}{1-\nu^2} h^3~. \label{energyEfrati1}
\end{align}

Now, take plate $s$ and deform it into a shape of a hemisphere with a radius $\frac{1}{2}\sqrt{2}r$ (an area preserving deformation). Following the authors' publication, one finds that the mid-surface can be express by the following map, $$\boldsymbol R (x^1,x^2) = \frac{1}{2}\sqrt{2} r \boldsymbol(\sin(x^1) \cos(x^2), ~\sin(x^1)\sin(x^2), ~\cos(x^1)\boldsymbol)_\text{E}$$ (see the definition of $\boldsymbol R(r,\theta)$ from section 4.1 of Efrati \emph{et al.} \cite{ efrati2009elastic}), and as one knows the deformed configuration in advance, one finds that the reference metric has the form $\bar{\boldsymbol g} (x^1,x^2) = \mathrm{diag}(\frac{1}{2} r^2, ~\frac{1}{2}r^2\sin^2(x^1),~1)$ (see equation 4.1 and the definition of $\Phi(r)$ from section 4.2 of Efrati \emph{et al.} \cite{ efrati2009elastic}), where this configuration is defined as the `stretch-free configuration' (see section 4.1 of Efrati \emph{et al.} \cite{ efrati2009elastic}). Thus, the stored energy of a circular plate that is being deformed into a hemisphere can be expressed as follows,
\begin{align}
E_s &=\frac{Y}{1-\nu}\int_{-\frac{1}{2}h}^{\frac{1}{2}h}\int_{0}^{\frac12 \pi} \int_{-\pi}^\pi \left(\frac{\sqrt 2 ~x^3}{r}\right)^2 \left(\frac{\sqrt 2~r}{2}\right)\sin(x^2) ~dx^1 dx^2 dx^3 \nonumber\\
& = \frac{1}{3}\pi \frac{Y}{1-\nu} h^3 ~. \label{energyEfrati2}
\end{align}

Equations (\ref{energyEfrati1}) and (\ref{energyEfrati2}), therefore, imply that, if one deforms a circular plate into a semi-cylinder with a radius $\frac{2}{\pi} r$ and deform a circular plate into a hemisphere with radius $\frac{1}{2}\sqrt{2} r$, then one gets the very similar respective energy densities $\frac{1}{96} \pi^2(1-\nu^2)^{-1}Y \varepsilon^2 $J$\text{m}^{-3}$ and $\frac{1}{3} Y(1-\nu)^{-1} \varepsilon^2 $J$\text{m}^{-3}$, i.e. both deformations' internal energies are of $\mathcal{O}(\varepsilon^2)$J$\text{m}^{-3}$. Which in turn implies that both deformations require force of $\mathcal{O}(\varepsilon)$N, given that one is applying the forces to the boundaries of the each respective plates. Thus, the authors' work asserts that it take approximately the same amount of force to bend a plate into a semi-cylindrical shape or to stretch a plate into a hemispherical shape with a similar radius. The reader may try this one's self: find a piece of aluminium foil (i.e. kitchen foil) and try to bend it over one's water bottle. This is a very simple process and the reader will able to accomplish this with a minimum of effort. In fact, the force of gravity is alone may even be sufficient to deform the piece of aluminium foil over the bottle without much interference. Now, try to stretch that same piece of aluminium foil smoothly over a rigid sphere with a similar radius, e.g. over a cricket ball. Can the reader do this without tearing or crumpling, and with the same force as one applied in the previous case?\\

To attempt this problem with mathematical regiour, consider the set $\mathscr{S} =\{\boldsymbol(x,y,0\boldsymbol)_\text{E} \in\textbf{E}^3 \mid x^2 + y^2 \leq r^2 \}$, which describes the mid-plane of the unstrained plates $c$ and $s$. Now, if one deforms plate $c$ is into a semi-cylindrical shape with a radius $\frac{2}{\pi} r$, then one finds that the map of the deformed  mid-surface has the following form, $$\boldsymbol R (x,y) = \boldsymbol(x, ~\frac{2}{\pi} r\sin\left(\frac{\pi}{2r} y\right),  ~\frac{2}{\pi} r\cos\left(\frac{\pi}{2r} y\right)\boldsymbol)_\text{E}~,$$ and thus, the total stored energy of a circular plate of radius $ r$ that is being deformed into a semi-cylindrical shape with a radius $\frac{2}{\pi} r$ can be expressed as follows,
\begin{align}
E_c & =\frac12 \frac{Y}{1-\nu^2}r\int_{-\frac{1}{2}h}^{\frac{1}{2}h}\iint_{\mathscr{S}} \left(\frac{\pi z}{2 r}\right)^2 dx dy dz \nonumber\\
& = \frac{1}{96}\pi^3 \frac{Y}{1-\nu^2} h^3~. \label{rightEnergy1}
\end{align}
Now, if one deforms plate $s$ is in to a hemisphere with a radius $\frac{1}{2}\sqrt{2} r$, then one finds that the map of the deformed mid-surface has the following form,
\begin{align*}
\boldsymbol R (x,y) = \frac{1}{2}\sqrt{2} \boldsymbol(x\sin\left(\frac{\pi}{2r} \sqrt{x^2+y^2}\right), ~y\sin\left(\frac{\pi}{2r}\sqrt{x^2+y^2}\right),&\\
 r\cos\left(\frac{\pi}{2r}\sqrt{x^2+y^2}\right)&\boldsymbol)_\text{E}~,
\end{align*}
and thus, the total stored energy of a circular plate with a radius $r$ that is being deformed into a hemisphere with a radius $\frac{1}{2}\sqrt{2} r$ can be expressed as follows
\begin{align}
E_s & =\frac12 \int_{-\frac{1}{2}h}^{\frac{1}{2}h}\iint_{\mathscr{S}} A^{\alpha \beta \gamma \delta}\epsilon_{\alpha\beta}(x,y)\epsilon_{\gamma\delta}(x,y) ~ dx dy dz + \mathcal{O}(h^3) \nonumber\\
& =C hYr^2 + \mathcal{O}(h^3) ~, \label{rightEnergy2}
\end{align}
where $C$ is an order-one positive constant that is independent of $h$, $Y$ and $r$, and
\begin{align*}
\boldsymbol{\epsilon} (x,y, 0)  = ~& \left(\frac r2\right)^2\frac{ \sin^2\left(\frac{\pi}{2r}\sqrt{x^2+y^2}\right)}{(x^2+y^2)^2}
\begin{pmatrix}
y^2 & -xy \\
-xy & x^2
\end{pmatrix} \\
& + \left(\frac \pi{4}\right)^2\frac{ 1}{x^2+y^2}
\begin{pmatrix}
x^2 & xy \\
xy & y^2
\end{pmatrix} -\frac12
\begin{pmatrix}
1 & 0 \\
0 & 1 
\end{pmatrix}~,
\end{align*}
is the strain tensor of the plate at $z=0$. As the reader can see from equations (\ref{rightEnergy1}) and (\ref{rightEnergy2}) that if one deforms a circular plate into a semi-cylinder with a radius $ \frac{2}{\pi} r$ and deform a circular plate into a hemisphere with a radius $\frac{1}{2}\sqrt{2} r$, then one get the respective energy densities $\mathcal{O}(\varepsilon^2)$J$\text{m}^{-3}$ and $\mathcal{O}(1)$J$\text{m}^{-3}$. Thus, one can see that it takes significantly higher amount of energy to deform plate in to a hemisphere than to simply bend it in to a semi-cylinder, as the former deformation requires a significant amount of stretching and compression, while the latter requires no such in-plane deformations, which is far more realistic than results obtained by Efrati \emph{et al.}'s  approach \cite{efrati2009elastic}. Note that the both deformations conserve area.\\

As further analysis, consider the deformed plate $s$ in curvilinear coordinates $\boldsymbol(x^1,x^2\boldsymbol)$, where $0\leq x^1\leq\frac12\pi$ and $|x^2|\leq\pi$. Now, the first and the second fundamental form tensors of the deformed configuration can be expressed respectively as $\boldsymbol F_{\!\text{[I]}} (x^1,x^2) =\frac12r^2 \mathrm{diag}(1,~\sin^2(x^1))$ and $\boldsymbol F_{\!\text{[II]}} (x^1,x^2) =-\frac12\sqrt{2}r\mathrm{diag}(1, ~\sin^2(x^1))$. If one follows the authors' publication, then one finds that the reference metric tensor has the form  $$\bar{\boldsymbol g} (x^1,x^2) = \mathrm{diag}(\frac12r^2,~\frac12r^2\sin^2(x^1),~1)~,$$ and this can only be derived by doing the following, $$\bar g_{ij}(x^1,x^2) = \partial_i r_{\bar k}(x^1,x^2,x^3) \partial_j r^{\bar k}(x^1,x^2, x^3)|_{x^3=0}~,$$ where $\boldsymbol r (x^1,x^2) = (\frac{1}{2}\sqrt{2} r+x^3) \boldsymbol(\sin(x^1) \cos(x^2), ~\sin(x^1)\sin(x^2), ~\cos(x^1)\boldsymbol)_\text{E}$. This implies that $\bar{\boldsymbol g} $ is the reference metric of a shell at $x^3=0$, and thus, $\bar{\boldsymbol g} $ is clearly not immersible in $\textbf{E}^3$ as  Ricci tensor is not identically zero, i.e. $$\boldsymbol {Ric} =\frac12r^2 \mathrm{diag}(1,~\sin^2(x^1))\neq\boldsymbol 0~, ~\text{for}~x^1\neq0 ~.$$ Thus, the authors' erroneous reference metric implies that $a_{\alpha\beta}-\bar g_{\alpha\beta} =0$, $\forall~\alpha,\beta\in\{1,2\}$, i.e. zero-planar strain, which in turn implies the existence of a `stretch-free configuration' for a substantially deformed plate.\\

Now, if one attempts this same problem with mathematical precision, then one finds that the reference metric tensor can be expressed as follows, $$\bar{\boldsymbol g} (x^1,x^2) = 4 \left(\frac{r}{\pi}\right)^2\mathrm{diag}(1 ,~(x^1)^2)~,$$ where $\bar g_{\alpha\beta} (x^1,x^2) = \partial_\alpha x\partial_\beta x +\partial_\alpha y \partial_\beta y$ with $x^1 = \frac{1}{2} r^{-1}\pi\sqrt{x^2+y^2}$ and $x^2 = \arctan(y/x)$. The coordinate transform $\boldsymbol(x(x^1,x^2), ~y(x^1,x^2)\boldsymbol)_\text{E}:\mathbb{R}^2 \to \textbf{E}^2$ is  a diffeomorphism (except at $x^1=x^2=0$) and Ricci tensor is identically zero, i.e. $\boldsymbol {Ric} =\boldsymbol 0$. Furthermore, $\det(\partial_1 x, \partial_2 x;\partial_1 y,\partial_2 y)> 0$, and thus, the definition of the unit normal to the deformed surface is not violated (again, except at $x^1=x^2=0$). Thus, half of change in the first fundamental form tensor (i.e. planar strain) can be expressed  as follows, $$\boldsymbol a(x^1,x^2) = \frac14 r^2 \mathrm{diag}(1 - \frac{8}{\pi^2},~ \sin^2(x^1) - \frac{8}{\pi^2} (x^1)^2)$$ and the change in second fundamental form tensor (i.e. bending) can be expressed as follows, $$\boldsymbol b(x^1,x^2) = -\frac12\sqrt{2}r\mathrm{diag}(1, ~\sin^2(x^1))~.$$ Now, with this coordinate transform no such `stretch-free configuration' can exist for a plate with a radius $r$ that is being deformed into a hemisphere with a $ \frac{1}{2}\sqrt{2} r$, unless the radius of the plate is zero.\\

Above analysis shows that Efrati \emph{et al.} \cite{ efrati2009elastic} are not studying plates, but they are studying nonlinear Koiter shells with an erroneous strain tensor. The authors' definition of the strain tensor leads to an incorrect change in second fundamental form tensor, and thus, an overestimation of the bending energy density of the shell per $h^3$ (see equation (\ref{efratiWrong})). To attempt this problem with mathematical precision, let $\tilde{g}_{ij}(\boldsymbol x) = \partial_i X_{\bar k}\partial_j X^{\bar k}$ be the metric of the reference configuration $$\boldsymbol X(\boldsymbol x) = \boldsymbol \sigma(x^1,x^2)+ x^3\frac{(\boldsymbol \sigma_{,1}\times\boldsymbol \sigma_{,2})}{||\boldsymbol \sigma_{,1}\times\boldsymbol \sigma_{,2}||}$$ with respect to the curvilinear coordinate system $\boldsymbol x = \boldsymbol(x^1,x^2,x^3\boldsymbol)$, where $\boldsymbol \sigma : \mathbb{R}^2 \to \textbf{E}^3$ is a sufficiently differentiable immersion (Efrati \emph{et al.}'s  reference metric \cite{efrati2009elastic} is derived by $\bar {\boldsymbol g} = \tilde{\boldsymbol g}|_{x^3=0}$). Thus, in nonlinear shell theory, one defines the strain tensor as $ \epsilon_{\alpha \beta}(\boldsymbol u) = \frac{1}{2} (g_{\alpha\beta}(\boldsymbol u) - \tilde{g}_{\alpha\beta})$, where $g_{\alpha\beta}(\boldsymbol u) =\partial_\alpha r_{\bar i}(\boldsymbol u)\partial_\beta r^{\bar i}(\boldsymbol u)$, $$\boldsymbol r(\boldsymbol u) =  \boldsymbol R (\boldsymbol u) + x^3 \frac{(\partial_1\boldsymbol R (\boldsymbol u)\times \partial_2 \boldsymbol R (\boldsymbol u))}{||\partial_1\boldsymbol R (\boldsymbol u)\times \partial_2 \boldsymbol R (\boldsymbol u)||} ~,$$ $\boldsymbol R (\boldsymbol u) = \boldsymbol \sigma + u^\alpha\partial_\alpha \boldsymbol \sigma + u^3\boldsymbol N $  and $\boldsymbol u(\boldsymbol x)$ is the displacement field in curvilinear coordinates. For more on nonlinear Koiter's shells, please consult Ciarlet \cite{Ciarlet}, Koiter \cite{koiter1966nonlinear}, and Libai and Simmonds \cite{libai2005nonlinear}.\\

Even if Efrati \emph{et al.} \cite{efrati2009elastic} obtain the correct form of the strain tensor for shells, they are still unjustified in using the shell strain tensor to model plates. To explain this matter with mathematical rigour, let $\mathscr{S}$ be a two-dimensional plane and let $\mathscr{S}^\prime$ be a two-dimensional surface. What the authors fail to grasp is that an arbitrary mapping from $\mathscr{S}$ to $\mathscr{S}^\prime$ (i.e. $\boldsymbol \sigma : \mathscr{S} \subset \mathbb R^2\to \mathscr{S}^\prime \subset\textbf{E}^3 $) is not the same as deforming the plane $\mathscr{S}$ into the surface $\mathscr{S}^\prime$ (i.e. $\{\mathscr{S}\cup\{\boldsymbol u\in\textbf{E}^3\}\subset\textbf{E}^3 \}= \{\mathscr{S}^\prime \subset \textbf{E}^3 \}$). The former is a simple coordinate transform (which may or may not be related to deforming the body), while the latter is a unique vector displacement (unique up to a rigid motion). To understand the distinction between a coordinate transform and a vector displacement, please consult section 1 and section 2 of Morassi and Paroni \cite{Morassi}.\\

\section{Conclusions}

In conclusion, Efrati \emph{et al.}'s \cite{efrati2009elastic} publication is not on plate theory: it is on shell theory with an incorrect strain tensor. We showed that the authors made the fundamental error of assuming that an arbitrary mapping from a plane to a curved surface is the same as deforming a plane in to a curved surface, when deriving their model. Thus, the authors numerical results imply that a thin object can be stretched substantially with very little force, which is physically unrealistic and mathematically disprovable.  All the theoretical work of the authors, i.e. nonlinear plate equations in curvilinear coordinates, can easily be rectified with the inclusion of both a sufficiently differentiable $\mathbb{R}^2 \to \textbf{E}^2$ diffeomorphism and a set of external loadings, such as an external strain field.

\bibliographystyle{./model1-num-names}
\bibliography{CriticaStudyOfEfratisWork}%
\biboptions{sort&compress}

\end{document}